\documentclass[fleqn,usenatbib]{mnras}
\usepackage{newtxtext,newtxmath}
\usepackage[T1]{fontenc}

\DeclareRobustCommand{\VAN}[3]{#2}
\let\VANthebibliography\thebibliography
\def\thebibliography{\DeclareRobustCommand{\VAN}[3]{##3}\VANthebibliography}

\usepackage{graphicx}	
\usepackage{amsmath}	
\usepackage{multirow}

\newcommand{\kepler}{\textit{Kepler}}
\newcommand{\ktwo}{\textit{K2}}
\newcommand{\gaia}{\textit{Gaia}}
\newcommand{\logg}{$\log g$}
\newcommand{\teff}{$\mathrm{T_{eff}}$}
\newcommand{\emcee}{{\scshape emcee}}
\newcommand{\totalflares}{\textcolor{black}{4430}}
\newcommand{\totalstars}{\textcolor{black}{403}}
\newcommand{\shiftstars}{\textcolor{black}{26}}

\newcommand{\origshiftflares}{\textcolor{black}{515}}
\newcommand{\tess}{\textit{TESS}}
\newcommand{\minenergyvalue}{\textcolor{black}{$3.7\times10^{29}$}}
\newcommand{\maxenergyvalue}{\textcolor{black}{$1.5\times10^{35}$}}
\newcommand{\kms}{km\,s$^{-1}$}

\title[Flares from blended and neighbouring \kepler\ stars]{Stellar flares from blended and neighbouring stars in \kepler\ short cadence observations}

\author[J. A. G. Jackman et al.]{
James A. G. Jackman,$^{1, 2, 3}$\thanks{E-mail: \textcolor{black}{jamesjackman@asu.edu}},
Evgenya Shkolnik$^{1}$, R. O. Parke Loyd$^{1}$
\\
$^{1}$School of Earth and Space Exploration, Arizona State University, Tempe, AZ 85287\\
$^{2}$Dept. of Physics, University of Warwick, Gibbet Hill Road, Coventry CV4 7AL, UK\\
$^{3}$Centre for Exoplanets and Habitability, University of Warwick, Gibbet Hill Road, Coventry CV4 7AL, UK\\
}

\date{Accepted XXX. Received YYY; in original form ZZZ}

\pubyear{2021}

\begin{document}
\label{firstpage}
\pagerange{\pageref{firstpage}--\pageref{lastpage}}
\maketitle

\begin{abstract}
We present the results of a search for stellar flares from stars neighbouring the target sources in the \kepler\ short cadence data. These flares have been discarded as contaminants in previous surveys and therefore provide an unexplored resource of flare events, in particular high energy events from faint stars. We have measured M dwarf flare energies up to \maxenergyvalue\ erg, \textcolor{black}{pushing the limit for flare energies measured }using \kepler\ data. We have used our sample to study the flaring activity of wide binaries, finding that the lower mass counterpart in a wide binary flares more often at a given energy. Of the 4430 flares detected in our original search, 298 came from a neighbouring star, a rate of $6.7\pm0.4$ per cent for the \kepler\ short cadence lightcurves. We have use our sample to estimate a $5.8\pm0.1$ per cent rate of false positive flare events in studies using \tess\ short cadence data.
\end{abstract}


\section{Introduction} \label{sec:introduction}

Blending of multiple stars within apertures is a known problem in astronomy. Additional light from neighbouring stars will dilute astrophysical signals from a target star of interest, or even introduce signals where there were none before. A noted example of this is in exoplanet transit searches, where a deep signal from a faint eclipsing binary within a photometric aperture is diluted by the brighter target star, resulting in an apparent transit signal. False positive signals such as this, if not fully understood, can impact large scale statistical studies, such as the incidence of stars hosting a transiting planet. Vital tools in the analysis of astrophysical signals are the image centroid and the use of difference images. Measuring the  position of the centroid, the centre of light, during the transit can identify a neighbouring star as the true source which can be flagged for removal \citep[e.g][]{Bryson14,Gunther17}. 

Another area of astronomy where blending can be an issue is the study of stellar flares. Stellar flares can be observed across a wide range of wavelengths, from the radio, through the infrared and optical up to X-rays \citep[e.g.][]{Jackson89,Flaccomio18}. Believed to be analogous to Solar flares, stellar flares are explosive phenomena which are the result of magnetic reconnection events in the upper atmospheres of stars \citep[e.g.][]{Benz10}. Charged particles are accelerated downwards from the reconnection site in the corona, spiralling along the newly reconnected magnetic field lines until they reach the dense chromosphere. There they are suddenly decelerated, evaporating the surrounding plasma and heating local atmospheric layers, which in turn results in white-light emission \citep[e.g.][]{Antonucci83,Hawley92,Fletcher08}. 

In recent years it is the white-light flare emission that has been the focus of ardent research. This is due to their detectability in wide-field optical exoplanet and transient surveys, which survey large patches of sky for long durations \citep[e.g.][]{Walkowicz11,Davenport16,Lin19, Schmidt19}. Such surveys have been used to study white-light flare energies, occurrence rates and their incidence rates. This has been both for individual stars and as a function of spectral type and age \citep[][]{Chang15,Ilin19,Jackman20,Feinstein20}. The role of stellar flares in exoplanet habitability has also been called into question as well, with the high energy UV emission from flares potentially altering the chemistry and structure of exoplanet atmospheres \citep[e.g.][]{Venot16,Chadney17}. At the same time, stellar flares have been suggested in the literature as a potential requirement for abiogenesis around M stars \citep[e.g.][]{Rimmer18}. In this scenario, flares provide the near-UV flux \citep[$\approx$ 200-300\,nm;][]{Ranjan17} needed for the formation of amino acids which may lead to abiogenesis. Relative to solar-type stars, the near-UV flux from main sequence M stars is lacking and as such flares may be a requirement to provide needed UV flux. Large scale flare studies \citep[such as those performed with \kepler\ and \tess, e.g.][]{Gunther2020} have aimed to understand the role of flares on the habitability of exoplanets around M stars by measuring the energies and flare occurrence rates for samples of M dwarfs and comparing these values with the predictions of atmospheric and biological models \citep[e.g][]{Tilley19}. However, these properties can be affected by blended flare events (e.g. by attributing M star flares to a nearby brighter star), affecting the accuracy of the measured properties and any comparisons to theoretical models. Consequently, the rate of blended flare events in such surveys requires careful characterisation so that surveys can ensure more accurate measurements of flare energies and occurrence rates. 

Previous large scale flare studies using \kepler\ observations have noted the issue of blending and taken various steps to get around it through a combination of vetting their input catalog and manual validation of flare events. For example, in their study of flares from G stars with \kepler\ long cadence observations, \citet{Shibayama13} only studied stars in their sample which did not have another star (as specified from the \kepler\ input catalog) within 12\arcsec, to limit contamination from close neighbours. Flare candidates were also vetted to remove events where the spatial brightness distribution of the target between the flare and quiescence (i.e., the flare comes from a star elsewhere in the aperture). Similar approaches have been used in subsequent studies \citep[e.g.][]{Yang17,Yang18}, although some have also added a step to remove eclipsing binaries from their search. \citet{Shibayama13} found that approximately 10 per cent of flares in their long cadence sample were due to a neighbouring star in their aperture. Similar values have been reported by other surveys \citep{Gao16,Yang17,Yang19}. However, by limiting their input samples before searching for flares, these studies did not measure the true rate of blended flare events, instead the rate from stars greater than some distance (typically 12\arcsec) from the host star. For studies which cannot spatially resolve stars down to this limit (i.e. those using \tess\ data) these previously reported values may underestimate the true false positive fractions of flares from neighbouring stars. 

Through their vetting, these studies removed thousands of flare events from neighbouring stars \citep[e.g.][]{Yang17}. These events are a currently unexplored resource and as such could provide an opportunity to measure some of the highest energy flares. Considering the case of a faint star nearby a brighter target star, only the highest energy events from the fainter star will contribute a detectable level of flux to the combined source. Studies using a similar technique of specifically targeting faint stars alone with wide-field surveys and waiting for them to flare into view \citep[e.g.][]{Schmidt16,Jackman19, Schmidt19} have been able to detect some of the highest energy flares from M and L dwarfs, with \citet{Schmidt19} measuring V band flare energies up to $10^{35}$ erg from M dwarfs observed with ASAS-SN. Consequently, blended flares, specifically if they come from faint sources, \textcolor{black}{can} provide a way of studying higher energy flare events than those typically observed in standard large scale studies.

In this work we present the results of a search for blended flares from \kepler\ short cadence observations. These flares have been missed or discarded by previous flare surveys, making this a trove of new events. We will discuss our method for detrending the \kepler\ short cadence observations, detecting flare events and then identifying those which are from stars neighbouring the target source.  We will detail our method for determining stellar properties and measuring flare amplitudes and energies. We will then discuss how these flares compare to those previously observed with \kepler\ and other surveys, along with measuring the fraction of flares which were blended in the \kepler\ short cadence observations.


\section{Data}
The primary \kepler\ mission began in 2009 and ran for nearly four years, observing a 115 deg$^{2}$ patch of sky in the northern hemisphere \citep[][]{Borucki10}. During this time, \kepler\ observed $\sim$ \textcolor{black}{115,000} stars \citep[][]{Batalha10}. The \kepler\ photometry was measured using two time cadences, a $\approx$ 30 minute long cadence mode and a $\approx$ 1 minute short cadence mode. The long cadence observations were performed for each target almost continuously throughout the duration of the mission. These long baseline lightcurves have been used for a variety of astrophysical studies, including but not limited to asteroseismology, stellar rotation and studies of eclipsing binaries and exoplanets \citep[e.g.][]{Bedding10,McQuillan14,Slawson11}. The short cadence observations were obtained for a small fraction of targets \citep[0.3\% in Q0;][]{Gilliland10} and for a fraction of the mission duration. The short cadence observations enabled studies of variability which were not possible with the long cadence mode, in particular the detection of short duration stellar flares. 

\kepler\ did not obtain photometry for every single star within its field of view, instead targets were pre-selected \citep[e.g.][]{Batalha10}. A ``postage stamp'' of pixels was selected for each target per quarter, within which the data was collected and stored. Each postage stamp was selected to include the flux from the star and surrounding background. The size of each postage stamp varied with the magnitude of each target, with brighter stars requiring larger stamps. \textcolor{black}{The average size of a postage stamp in the full set of short cadence observations is $12\times8$ pixels, approximately $48\arcsec\times32\arcsec$}. The postage stamp pixel files were downloaded at the end of each \kepler\ quarter and made available for download as ``Target Pixel Files''. These target pixel files include the individual pixel images, timestamps, quality flags, row and column image centroids and the optimal aperture. Lightcurves and ancillary products (e.g. image centroids) for each target were automatically generated from these target pixel files using the optimal apertures, which were designed to maximise the signal-to-noise ratio \citep[][]{Haas10}.

As each postage stamp includes some amount of sky around the target star, there is a chance nearby sources will have been included in the resultant target pixel files. If these stars are close enough to the target star and they flare with enough energy, then this flare will be visible in the \kepler\ lightcurve. By inspecting the individual images and the image centroids we can identify these events and use them to characterise previously unexplored stellar flares.

\section{Methods}

\subsection{Detrending and Flare Detection}
In this work we used the Simple Aperture Photometry (SAP) data for all stars with short cadence observations. Before searching for flares we detrended the SAP lightcurves. This was done to remove instrumental trends and astrophysical variability (e.g. due to starspots) which may hinder accurate flare detections. Our detrending process is similar to those found in previous flare studies, in particular \citet{Yang19}, and is as follows. 

For each target lightcurve, we initially masked out all data points with quality bit flags equal to 1, 2, 3, 4, 5, 6, 7, 9, 13, 15, 16 and 17. These are the recommended quality flags masked by the \kepler\ data reduction pipeline and a flag for reaction wheel zero crossing \citep[][]{keplerarchive}. Before smoothing a lightcurve using a median filter, we checked whether the target source had a known rotation period, using the catalogs of \citet{Reinhold13} and \citet{McQuillan14}. This was to to ensure the selected window could effectively smooth out any rotational modulation present. If the source did not have a known rotation period, we searched for the strongest period with a generalised Lomb Scargle periodogram, using the astropy LombScargle routines \citep[][]{astropy13}. We sampled each lightcurve with periods between 100 minutes and half the duration of the full lightcurve. If the best period from the generalised periodogram had a power greater than 0.25 then this period was chosen. This power threshold was chosen empirically to avoid false positive periods in our dataset \citep[e.g.][]{Oelkers18}. After this routine we smoothed the lightcurve.

Before smoothing we initially split the lightcurve into separate continuous segments, using gaps in coverage greater than 30 minutes to split the data. This was done to avoid jumps in the measured flux between quarters, or segments with quarters, affecting our smoothing algorithm. Continuous segments less than one quarter of a day were masked. For each continuous segment we applied a median filter. The window size was selected to be equal to one tenth the measured period, with lower and upper limits of 30 minutes and 12 hours respectively. These limits were chosen to make sure lightcurves were not excessively smoothed at the lower level (in turn removing detected flares) or suffered from overly large window sizes. For sources without measured periods we used a standard window size of 100 minutes, which was used by \citet{Pugh16} to remove lightcurve modulation without removing flares. Each segment was then iteratively smoothed. In each iteration, the original lightcurve was divided by a smoothed version. Outlying points greater than 3$\sigma$ from the median of the resultant lightcurve were then masked. The equivalent points were masked in the original lightcurve and were interpolated over. This process was repeated up to 20 times, or until there were no more outlying points. Once the process was completed for a given continuous segment, the original segment (which contains the flares and rotational modulation) was divided by the final smoothed version (which should only contain the rotational modulation), to give what should be a flare-only segment. We then searched for flares in this detrended segment. 
To identify flares in the detrended lightcurve segment, we searched for consecutive outliers 3$\sigma$ above the lightcurve median. Here $\sigma$ refers to the standard deviation of the detrended lightcurve. Regions with more at least three consecutive outliers were automatically flagged as flare candidates. The times at which the lightcurve first goes above and below the 3$\sigma$ limit are recorded as the flare start and end time. Flare candidates starting and ending within 2.4 hours (0.1 days) of each other were combined to avoid multiple detections of the same event (e.g. due to a multi-peaked flare). Once all continuous segments had been run for a target, all flare candidates were combined and saved. 

Some sources in the short cadence data are known eclipsing binaries. Deep eclipses in the \kepler\ lightcurves, if not masked, can increase the measured variance. This in turn will increase the $3\sigma$ limit used for detecting flares, meaning only the highest amplitude and energy events will be detected. Due to the effects of dilution, flares from neighbouring sources are unlikely to have large amplitudes in the original \kepler\ lightcurves. If a target source was a known eclipsing binary and it is included in the \kepler\ eclipsing binary catalog\footnote{\url{http://keplerebs.villanova.edu/}} \citep[][]{Prsa11,Slawson11,Kirk16}, we masked its eclipse before smoothing the lightcurve and searching for flares. Eclipses were masked by iteratively applying a median filter to the lightcurve with a user-defined window of one fifth the orbital period, then masking outliers $3\sigma$ below the median of the lightcurve. These masked points were flagged and stored. The masked points were then interpolated over. This process was repeated either 20 times, or until no points were flagged as outliers. The stored masked points in the original lightcurve were then replaced using a cubic spline fit using anchor points two and one hour either side of the masked eclipses. This new lightcurve, with the eclipses replaced by a smoothed spline, was then put forward for the iterative smoothing and automated flare detection.

Flare candidates flagged by our automated method were then vetted to determine their authenticity. To vet flare candidates we first flagged candidates which had the same start and end times, as determined from the \kepler\ BJD time stamps. Stellar flares are known to be stochastic events, meaning any candidate events starting and ending at the same times on different stars are more likely to be instrumental than astrophysical in origin. Remaining flare candidates were visually inspected to remove those due to other sources of astrophysical variability \citep[e.g. RR Lyrae or high frequency pulsators, e.g.][]{Yang19}, or those which may have been flagged as a result of our detrending algorithm. 

\subsubsection{Centroiding and Image Analysis} \label{sec:cen_im_anal}
\begin{figure*}
    \centering
    \includegraphics[width=\textwidth]{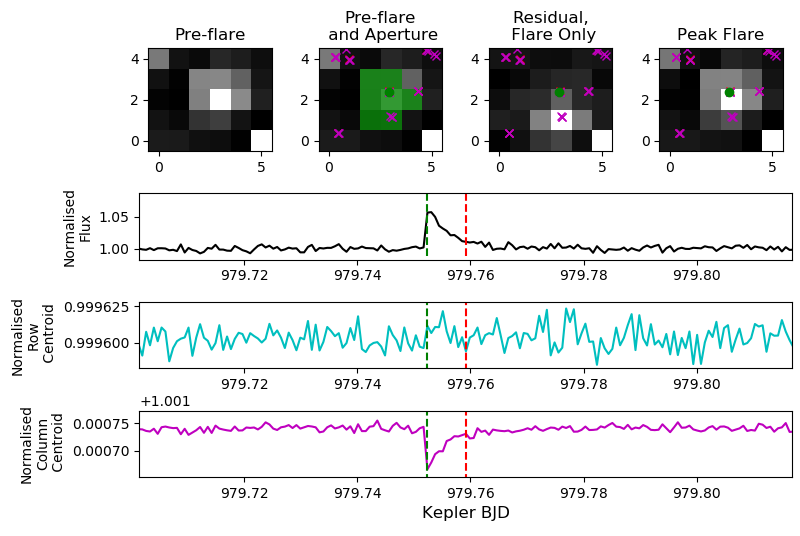}
    \caption{Example of the flare image and centroiding analysis, for a flare from the M3 dwarf KIS J192818.57+383757.2. The top panel shows the pre-flare, residual and peak flare images. The green squares indicate the position of the original \kepler\ aperture. The green circle is the position of the target source in the \kepler\ input catalog. The purple crosses indicate the positions of stars in the Kepler-INT survey. The bottom three panels show the detrended lightcurve and the positions of the row and the column centroid. Note the clear shift in centroid position during the flare, showing it did not come from the target source.}
    \label{fig:flare_image_shift}
\end{figure*}

After vetting all automatically flagged candidates to obtain a clean sample, we inspected the centroids and images of all our flares. The purpose of this was to identify which flares arise from stars neighbouring the target stars. For all events we obtained a pre-flare image from two hours before the start of the flare and the image at the peak of the flare. We subtracted the pre-flare image from the image at the flare peak to obtain a residual, ``flare only'', image. We compared the pre-flare, residual flare only and original flare images for every event. Flares from stars neighbouring the target source will have different source locations in the pre-flare and residual images. An example is shown in Fig.\,\ref{fig:flare_image_shift}. We identified candidate flares from nearby stars through visual inspection of the residual images. This method of comparing images was used in \citet{Yang17} to identify possible blends and flares from nearby sources. 
We also used the centroid position as an extra confirmation that flares came from a nearby object. Flares which come from stars other than the target star will shift the centre of light of the aperture towards the true flare source. 

For all sources we inspected whether the pipeline aperture fully encompassed the PSF of the flaring source. In some cases the flare source lay in a pixel adjacent to the edge of the pipeline aperture. In these cases, light from the flare would enter the aperture of the target source, resulting in the appearance of a small flare event. Where required we created new apertures by eye, designed to cover the PSF of the flaring source. All the new apertures were made to include the light from both the target and neighbouring star, due to the separations not being large enough to fully disentangle the PSFs of the two stars. We used the new apertures to generate improved lightcurves directly from the target pixel files, upon which our flare detection and verification algorithms were run again to find previously undetected flares. 

\subsection{Determining the
contribution of all stars} \label{sec:dilution}
For apertures covering multiple stars it was important to determine the relative flux contribution from each source. Flux from nearby sources will dilute flare lightcurves and if not corrected for, this results in reduced measured flare amplitudes and energies. To determine the flux contribution from sources within an aperture we calculated their \kepler\ magnitudes. As some sources in our apertures did not have entries in the \kepler\ input catalog, we decided to calculate the \kepler\ magnitudes of each source ourselves. 
Where possible, we calculated this by applying an offset to their catalogue \gaia\ DR2 \textit{G} magnitude. The offset between \gaia\ \textit{G} and the \kepler\ magnitude was determined by crossmatching all sources within the \kepler\ input catalog with those in \gaia\ DR2 which had passed the recommended photometric and astrometric quality checks \citep[][]{Arenou18, Lindegren18}. When sources did not have a \gaia\ \textit{G} band magnitude, but did have an entry in the \kepler-INT catalog, we used the available INT $g$, $r$ and $i$ magnitudes with Eq.\,(2)-(5) from \citet{Brown11} to estimate the \kepler\ magnitude. \textcolor{black}{The dilution is calculated as the ratio of the flux from the flaring stars to the flux from all stars within the aperture.}

To obtain broadband photometry for all our identified flare sources, we performed crossmatching with various photometric catalogues. We crossmatched each source with \gaia\ DR2 \citep[][]{Gaia18}, APASS \citep[][]{APASS_14},
2MASS \citep[][]{2MASS_2006}, WISE \citep[][]{ALLWISE2014}, Pan-STARRS \citep[][]{Chmabers16} and the Kepler-INT survey. Sources which were unresolved in APASS, 2MASS or WISE, but are resolved in \gaia, were flagged for reference during our SED fitting.

\subsection{Stellar Properties} \label{sec:sed_fitting}

Where possible we fit the Spectral Energy Distribution (SED) of each of our flare stars using the PHOENIX v2 spectral library \citep[][]{Allard12}. In order to calculate flare energies, as we will discuss in Sect.\,\ref{sec:flare_energy_calculation}, we required a star's radius and effective temperature. For all stars we fitted for the effective temperature \teff, the surface gravity \logg, interstellar extinction $A_{V}$ and an uncertainty scale factor $\sigma$. We used the uncertainty scale factor to account for underestimated variability in reported catalogue photometry. When fitting for interstellar extinction we assumed average Milky Way parameters for the reddening and used $R_{V}=3.1$, where $R_{V}$ is the ratio of total to selective extinction \citep[][]{Cardelli89}. We used the 3D dust maps of \citet{Green19} to apply Gaussian priors on our fitted extinction values. These dust maps are based on \gaia\ DR2, Pan-STARRS and 2MASS and can be used to estimate the expected level of extinction along a given line-of-sight, both in total and as a function of distance. Each template SED was multiplied by a scale factor $S$, equal to $(R_{*}/d)^2$, where $R_{*}$ is the stellar radius and $d$ is the distance. For stars which had a distance measurement from \citet{BailerJones18}, we fit directly for the stellar radius $R_{*}$ and the distance $d$. We applied a Gaussian prior to the distance, using the values from \citet{BailerJones18}.  We included the distance in our fitting to propagate its uncertainty to our fitted radius. For stars without a distance measurement, we fit the scale factor only. To determine the radii of these objects, we assumed the star was on the main sequence and calculated the radius using the temperature-radius relations of \citet{Boyajian12,Boyajian17}.

Some sources were too close to a nearby, brighter, neighbour to have non-blended photometry in all of our crossmatched catalogues. In these cases, where possible, we fit the SED of both the flaring star and the nearby neighbour simultaneously, following a method similar to that used in \citet{Jackman19}. In this scenario we fit for the effective temperature, surface gravity, interstellar extinction and scale factor of both stars simultaneously, along with the uncertainty scale factor $\sigma$. We fit the combined SED of both stars to blended catalogue magnitudes and the individual SEDs to \gaia\ resolved catalogue photometry. This is primarily the \gaia\ $G$ band photometry, although depending on the separation can include Pan-STARRS.

To fully explore the posterior parameter space we used the \emcee\ Python package \citep[][]{emcee13} to generate a Markov Chain Monte-Carlo (MCMC) process. For all SED fits we ran the MCMC process for 10,000 steps using 32 walkers and discarded the first 3000 as a burn-in. Some stars did not have enough broadband photometry for SED fitting. For these stars we used the \teff\ and radii colour-colour relations from \citet{Stassun19} with the \gaia\ $G$, $RP$ and $BP$ photometry and the distance measurements from \citet{BailerJones18}. These relations were created for the TESS Input Catalog (TIC). 

\subsection{Calculating Flare Energies} \label{sec:flare_energy_calculation}
Before calculating the energy we normalised each flare lightcurve by dividing it by the quiescent flux value. We assumed that any stellar variation (e.g. due to rotational modulation) during the flare changed on timescales longer than the flare duration and as such could be fitted with a linear baseline, which was obtained by linearly interpolating the lightcurve just before and after each flare event. The flare lightcurve was then divided by this baseline to obtain the normalised flare lightcurve. The quiescent baseline of this normalised flare lightcurve was set to zero. To account for the possible effects of dilution from other stars in our apertures, we divided each normalised flare lightcurve by the dilution values calculated in Sect.\,\ref{sec:dilution}. This renormalised flare lightcurve was then used to measure the flare amplitude and calculate the flare energy. We express the flare amplitude using $\Delta F/F_{q}$ where $\Delta F$ is the flux due to the flare and $F_{q}$ is the quiescent flux of the star \citep[e.g.][]{Hawley14}. Using this formulation, a value of 0 indicates no flare is present, while a value of 1 indicates the flare emits as much flux (in the chosen filter) as the quiescent star, or a doubling of the total observed flux.

To calculate the energy of each flare we followed the method outlined by \citet{Shibayama13}. We assumed the flare spectrum can be modelled by a single blackbody with a temperature of $9000\pm500$\,K. Multi-colour photometric and spectroscopic flare observations \citep[][]{Hawley92,Kowalski13}  have shown that flares can be approximated as a 9000\,K blackbody. However, these and more recent studies have also shown that flares exhibit changing temperatures between individual events, with continuum blackbody temperatures up to 15,000 and in some cases even up to 40,000\,K being measured from optical and far ultraviolet observations of the impulsive phase of flares \citep[][]{Loyd18,Froning19,Howard20}. In addition to this, the flare temperature has been observed to change within events, cooling during the flare decay. \textcolor{black}{We calculated how the measured flare energy, using the \citet{Shibayama13} method with the \kepler\ bandpass, would change if we used temperatures of 6000\,K and 16000\,K, values representative of the minimum and maximum flare temperatures measured spectroscopically by \citet{Kowalski13}. We found that the measured energies would range from 0.8 to 2.5 times that the value calculated when a 9000\,K blackbody is assumed.} Consequently, the assumption of a 9000\,K blackbody is only an approximation to the true behaviour of white-light flares and may underestimate the energy, particularly during the flare rise \textcolor{black}{when the flare temperature reaches its maximum}. Studies apply a 500\,K error to partially account for the changing flare temperature, as the changing flare temperature behaviour cannot be accurately predicted. We note that this uncertainty may need to be revised in future, as more measurements of flare temperatures become available. However, as this method has been used several times in previous flare studies using \kepler\ and \tess\ \citep[e.g.][]{Shibayama13,Yang18,Gunther2020,Tu20} it allows for a comparison between our results and previous studies and as such we use it in this work.

\section{Results} \label{sec:results}
From our survey of the \kepler\ short cadence data we identified \totalflares\ flares from \totalstars\ stars. We identified that \origshiftflares\ flares in \shiftstars\ \kepler\ short cadence lightcurves were either due to a nearby source, or due to there being flares from both the target star and a close companion (likely excluding it from other surveys). 
From the \shiftstars\ \kepler\ lightcurves there were 34 individual flaring stars, 26 of which were neighbouring the target source. The remaining eight were cases when both stars in the postage stamp flared. 30 of the 34 stars had enough catalogue photometry for SED fitting or the \citet{Stassun19} colour-colour relations. Further inspection with revised apertures, as discussed in Sect.\,\ref{sec:cen_im_anal}, resulted in an extra 68 flare detections from nearby stars, increasing the total number of flares to 4498, with 583 flares from 34 stars. \textcolor{black}{A full table of our results is provided in the online version of this work.}

\subsection{Flare Amplitudes}
We measured the flare amplitudes using the method outlined in Sect.\,\ref{sec:flare_energy_calculation}, by accounting for the effects of dilution from nearby stars. 515 of the flares in the final sample of 583 flares were from M stars and 16 of these had corrected flare amplitudes greater than 1. Note that these 515 flares are not the full 515 originally detected using the standard \kepler\ apertures. The full sample of M dwarf flares have amplitudes between 0.002 and 35.7. The highest amplitude flare, from the M3 dwarf KIS J192818.57+383757.2 (PS1 154352920774099705), is equivalent to $\Delta \mathrm{m_{Kp}}$=3.9. Assuming a 9000\,K blackbody, we estimate that this would be equivalent to a 4.8 magnitude change in the V filter and 7.5 magnitudes in the U filter for that star. The original \kepler\ lightcurve for this flare is shown in Fig.\,\ref{fig:flare_image_shift}. This flare is also greater in amplitude than any of those studied from \ktwo\ observations of M dwarfs by \citet{Lin19} and those in the sample of M dwarf hyperflares studied using \kepler\ and LAMOST by \citet{Chang18}. However, while having higher amplitudes than many flares studied previously using \kepler\ or \ktwo\ data, the larger flares in our sample are comparable to those from other surveys which studied much greater numbers of M dwarfs, in particular EvryScope \citep[][]{Howard19}, ASASSN \citep[][]{Schmidt19,Rodriguez20} and \tess\ \citep[][]{Gunther2020}.

\subsection{Flare Energies}
\begin{figure}
    \centering
    \includegraphics[width=\columnwidth]{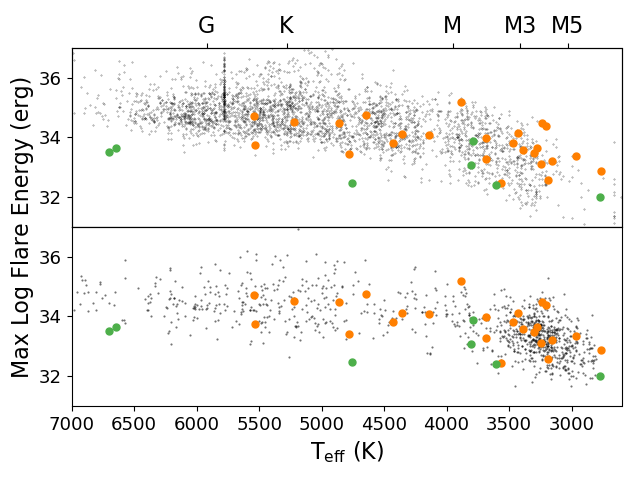}
    \caption{Top: Comparison of the maximum flare energies in our sample with those detected with \kepler\ long cadence observations from \citet{Yang19} (black points). The orange points are stars neighbouring the target sources. The green points are target sources which also flared in our observations. Bottom: The same, but now using flares detected with \tess\ 2 minute cadence observations by \citet{Gunther2020}. }
    \label{fig:max_energy}
\end{figure}

The flare energies were calculated following the method outlined in Sect.\,\ref{sec:flare_energy_calculation}. 
The flare energies from all stars in our sample \textcolor{black}{range from \minenergyvalue\ to \maxenergyvalue\ erg. The flare energies from the target sources sit within this range.}
The maximum flare energy as a function of spectral type, compared with the sample of flares detected from \kepler\ and \ktwo\ by \citet{Yang18} and the sample of flares detected with \tess\ short cadence observations by \citet{Gunther2020} is shown in Fig.\,\ref{fig:max_energy}. \textcolor{black}{We can see that the flares from neighbouring M stars in our sample have maximum energies comparable to those previously detected using \tess\ short cadence data, while some appear to push the limit for flares detected with the \kepler\ long cadence data. We believe this is due to the larger number of M dwarfs observed with \tess\ relative to \kepler, aiding the detection of high energy events.}

We believe that the reason why our M star energies are \textcolor{black}{preferentially} higher than previous studies with \kepler\ is due to a selection effect related to our detection and confirmation methods, which we will explain here. Due to their low luminosities, an M star neighbouring a brighter star (e.g. a G star) will not contribute much flux in the \kepler\ aperture. Therefore, only the highest energy flares, with the exact energy depending on the flux ratio between the two stars, will contribute enough flux to trigger our detection method and also cause an observable centroid shift. Small flares are less likely to trigger our detection method. As we go to higher luminosity stars, which contribute more quiescent flux, we can detect lower energy flares. This can be seen in Fig.\,\ref{fig:max_energy}, where our K stars broadly reside within the \citet{Yang19} \kepler\ long cadence sample. 

The maximum energies of the F, G and K spectral type stars are comparable to those measured in previous studies. We can also see some flares in Fig.\,\ref{fig:max_energy} that are at the lower limit of the \kepler\ sensitivity. These sources are the hotter, brighter, components of pairs of stars in which both components flare. Some of these sources are in wide binaries, which we will discuss in Sect.\,\ref{sec:flare_energy_calculation}, and are less active than their lower mass companions over the \kepler\ lifetime. Consequently, these stars flare less often and in a given time span appear to flare with a lower maximum energy.

\subsection{Flare Occurrence Rates}
Some stars in our sample flared multiple times, allowing us to measure the flare occurrence rate. Previous studies have shown that flares occur with a power-law-like distribution in energy, written as
\begin{equation} \label{eq:edit_powlaw}
    dN(E) \propto E^{-\alpha}dE
\end{equation}
where $N(E)$ is the number of flares in a given duration with energy $E$ and $\alpha$ is the power law index. The number of flares which occur with an energy greater than $E$ can then be written as
\begin{equation} \label{eq:ffd}
    \log{N(E_{fl}>E)} = C + \beta \log{E}
\end{equation}
where $C$ is a normalisation constant and $\beta = 1-\alpha$. To measure $C$ and $\alpha$ for stars in our sample, we fit the flare occurrence rates using the ``{\scshape powerlaw}'' Python package \citep[][]{Alstott14}. {\scshape powerlaw} is designed to fit to heavy-tailed distributions and has previously been used to fit occurrence rates for flares observed with \kepler\ \citep[e.g.][]{Lin19}. For each flaring star in our sample we used {\scshape powerlaw} to identify the energy above which flares were consistent with a power-law distribution. Flares with energies lower than this value were those affected by selection effects due to the sensitivity of our detection method. This effect is well documented in stellar flare literature \citep[e.g.][]{Gershberg72,Kunkel73,Hawley14} and results in a turnover in the observed occurrence rate at low energies. If a star had five or more flares above the limiting energy, then we fitted Eq.\,\ref{eq:ffd} to its measured occurrence rate using a least squares fitting. We have presented the best fitting values of $C$, $\beta$ and $\alpha$, along with the information about each star and the number of flares detected and fit to, in Tab.\,\ref{tab:occ_rate}. 

We can see in Tab.\,\ref{tab:occ_rate} that \textcolor{black}{many of the} stars in our sample have $\alpha>2$. A value of $\alpha$ greater than 2 indicates that the total flare energy distribution of a star is dominated by low energy flare events \citep[e.g.][]{Gudel03}. If they dominate the total energy distribution, then low energy events such as nanoflares have been suggested as being able to heat the observable X-ray corona \citep[][]{Parker88,Hudson91}. We note then that if our fitted power law distributions hold to nanoflare energies \citep[$\sim 10^{24}$ erg;][]{Parker88}, then flares may be a major contributor to the heating of the quiescent corona for some of the stars in our sample.

\begin{table*}
    \centering
    \begin{tabular}{c|c|c|c|c|c|c|c|}
    \hline
    Name & \multicolumn{1}{|p{1cm}|}{\centering \teff\ \\(K)} & $C$ & $\alpha$ & \multicolumn{1}{|p{1cm}|}{\centering  $E_{lim}$ \\(erg)} & \multicolumn{1}{|p{1cm}|}{\centering  $E_{max}$ \\(erg)}& $N_{flares,fit}$ & \multicolumn{1}{|p{2cm}|}{\centering Time for $10^{33}$\\ erg (days)} \tabularnewline \hline
Gaia DR2 2099133865420292480&$4353\pm_{-188}^{+214}$ &$41.4\pm9.0$&$2.26\pm0.27$&$2.5\times10^{33}$&$1.3\times10^{34}$&13&1.4 \tabularnewline
Gaia DR2 2076373356250706176&$3790\pm_{-243}^{+323}$ &$41.3\pm12.1$&$2.27\pm0.37$&$7.1\times10^{32}$&$7.7\times10^{33}$&11&5.5 \tabularnewline
Gaia DR2 2101855569017361280&$4142\pm_{-25}^{+26}$ &$15.5\pm9.6$&$1.50\pm0.29$&$5.6\times10^{32}$&$1.2\times10^{34}$&5&9.2 \tabularnewline
Gaia DR2 2104786042380468736&$3686\pm_{-122}^{+125}$ &$29.2\pm3.3$&$1.91\pm0.10$&$1.8\times10^{32}$&$1.9\times10^{33}$&21&5.9 \tabularnewline
Gaia DR2 2105165163433824000&$5544\pm_{-79}^{+115}$ &$26.6\pm11.3$&$1.81\pm0.33$&$6.4\times10^{33}$&$5.2\times10^{34}$&7&1.0 \tabularnewline
Gaia DR2 2105329991397882624&$2964\pm_{-20}^{+29}$ &$14.9\pm4.4$&$1.48\pm0.14$&$8.7\times10^{31}$&$2.3\times10^{33}$&9&10.1 \tabularnewline
Gaia DR2 2105930840143687040&$3275\pm_{-61}^{+91}$ &$29.0\pm6.6$&$1.94\pm0.20$&$6.1\times10^{32}$&$4.4\times10^{33}$&11&106 \tabularnewline
Gaia DR2 2126410068846203776&$5099\pm_{-60}^{+76}$ &$47.6\pm21.1$&$2.46\pm0.62$&$6.0\times10^{33}$&$1.7\times10^{34}$&6&2.7 \tabularnewline
Gaia DR2 2130461253797240704&$4424\pm_{-28}^{+29}$ &$45.6\pm11.2$&$2.39\pm0.34$&$1.0\times10^{33}$&$6.5\times10^{33}$&12&2.5 \tabularnewline
Gaia DR2 2131725447350716032$\mathrm{^{*,1}}$&$3602\pm_{-59}^{+62}$ &$42.6\pm22.9$&$2.39\pm0.72$&$6.7\times10^{31}$&$2.5\times10^{32}$&6&2000 \tabularnewline
Gaia DR2 2131725451649506944$\mathrm{^{*,1}}$&$3190\pm_{-22}^{+27}$ &$54.7\pm4.8$&$2.74\pm0.15$&$7.3\times10^{31}$&$3.7\times10^{32}$&31&735 \tabularnewline
Gaia DR2 2132768956904826624&$3431\pm_{-146}^{+166}$ &$50.1\pm4.8$&$2.58\pm0.15$&$2.3\times10^{32}$&$1.3\times10^{34}$&36&95 \tabularnewline

Gaia DR2 2079073928612819840$\mathrm{^{*,2}}$&$2776\pm_{-34}^{+35}$ &$51.0\pm4.9$&$2.68\pm0.16$&$8.4\times10^{30}$&$2.0\times10^{31}$$({\dagger})$&35&17400 \tabularnewline
Gaia DR2 2079073928612821760$\mathrm{^{*,2}}$&$2763\pm_{-33}^{+31}$ &$36.8\pm1.6$&$2.23\pm0.05$&$2.5\times10^{30}$&$1.6\times10^{31}$$({\dagger})$&66&4830 \tabularnewline

    \hline
    
    \end{tabular}
    \caption{Best fitting power law parameters for the flare occurrence rates of stars in our sample. $E_{lim}$ is the minimum energy limit used when fitting the power law distribution. $N_{flares,fit}$ is the number of flares used in the fitting. $E_{max}$ is the maximum flare energy. The last column gives the predicted waiting time to observe a flare with an energy of \textcolor{black}{$10^{33}$} erg or greater. The $e$ symbol indicates this is an extrapolation of our fitted power law. The * indicates that two stars are both components of the same wide binary. The ${\dagger}$ symbol indicates that these stars, in GJ 1245, had an upper energy limit of \textcolor{black}{$2\times10^{31}$} erg imposed when fitting.}
    \label{tab:occ_rate}
\end{table*}

\subsection{Wide Binaries}
After stars arrive on the zero-age main sequence and start to undergo magnetic braking, their flaring activity is expected to decrease. This behaviour has been observed in observations of open clusters \citep[][]{Ilin19}, which provide a snapshot of a specific time of stellar evolution. The change in flare activity is expected to have both a dependence on age and stellar mass, as solar-type stars for example spin down faster than their lower mass counterparts. Another way of comparing how flaring activity changes with mass and age is to use wide binaries. Wide binary systems are pairs of stars with similar proper motions and distances. Their similar kinematics means wide binaries are thought to have formed together and be coeval, making them excellent laboratories for testing age-activity relations \citep[e.g.][]{Gunning14}. 

We have investigated whether any of our flaring sources are components of wide binaries or common proper motion pairs. We crossmatched the position of each of the 34 individual flaring stars with \gaia\ DR2 to obtain the parallax and proper motion information. We crossmatched sources with total proper motions greater than 40\,mas\,$\mathrm{yr^{-1}}$ with the SUPERWIDE catalog \citep[][]{Hartman20} of wide binaries. 7 flaring stars in our sample were listed as members of wide binaries in the SUPERWIDE catalog. For sources with total proper motions less than 40\,mas\,$\mathrm{yr^{-1}}$, we checked to see if there were nearby sources with similar parallaxes and proper motions. For each candidate pair we calculated the difference in distance, their projected physical separation and the difference in their tangential velocities. 
For a pair to be considered a wide binary, we required their projected separation be less than 10000 AU, the difference in tangential velocity be less than 10\,\kms and the difference in their distances to be less than three times the uncertainty \citep[e.g.][]{Oh17,Hollands18,Hartman20}. 
From this analysis we identified a further 11 flaring stars in our sample as likely members of wide binaries, bringing the total to 18 flaring stars from 13 wide binary systems.

Flares from wide binaries have previously been studied by \citet{Clarke18}, who crossmatched the \citet{Davenport16} flare sample with the \citet{Janes17} sample of common proper motion wide binaries. 
They found that the majority of equal mass binaries in their sample displayed similar flare activity and noted that those that didn't may have been affected by tides. For a star to be included in the \citet{Davenport16} sample, it had to have had at least 100 flare candidate events, a filter imposed to rule out false positive events. 
This requirement consequently biased the sample to more active stars, limiting the sample of wide binaries studied by \citet{Clarke18}. In addition to this, we have explicitly searched for events which will have been mistaken for flares from a single star in previous surveys, decreasing the chance they would have been flagged as wide binary flaring systems in the past. As a result of these filters, none of the stars in our sample which we have identified as belonging to wide binaries were studied in the sample of \citet{Clarke18}.

Of the 13 wide binary systems in our sample, there were five pairs where we could detect flares from both components. Of these five, there were two pairs where we could fit the flare occurrence rates of both components. For the other three we could fit the flare occurrence rate of one star in the pair. Of the remaining wide binaries, we detected at least one flare from one component but were unable to fit the flare occurrence rate. 
One wide binary detected as part of our survey is the M5+M5 binary GJ 1245AB. The flaring activity of GJ 1245 was studied thoroughly by \citet{Lurie15}, who performed PSF fitting to separate out the flux from GJ 1245A and B. Consequently we have not performed an in-depth analysis of this system. 
\citet{Lurie15} identified that up to 25\% and 9\% of flares from GJ 1245A and B respectively may have exceeded the linearity regime in the \kepler\ CCD. In our sample this limit corresponds to an energy of approximately \textcolor{black}{$2\times10^{31}$} 
erg. We have used this energy as an upper limit when fitting the occurrence rates for GJ 1245A and B. We find that on average, GJ 1245A flares more than twice as often as GJ 1245B and that at a bolometric energy of \textcolor{black}{ $1\times10^{31}$} 
erg (assuming a 9000\,K blackbody) GJ 1245A flares once every \textcolor{black}{eight} days and GJ 1245B once every \textcolor{black}{17} days.

We find, for all binary pairs in our sample that, as expected, the lower temperature component flares more often than the higher temperature component. We have either inferred this from the absence of flares detected from the hotter component, or from measured flare occurrence rates. This result is in agreement with \citet{Clarke18}. For pairs with flares from both stars, we extrapolated the occurrence rate measured from the lower temperature component to the maximum energy detected from the higher temperature component to estimate a ratio of the flare occurrences. The flare energies, effective temperatures and the estimated ratios are given in Tab.\,\ref{tab:ratios}. 

\begin{table*}
    \centering
    \begin{tabular}{|c|c|c|c|c|c}
    \hline
    Source ID 1 & \teff$_{,1}$ (K) & Source ID 2 & \teff$_{,2}$ (K) & Flare Energy (erg) & Ratio of 1 to 2 \tabularnewline \hline
    Gaia DR2 2076373356250706176 & $3790\pm_{-243}^{+323}$ & Gaia DR2 2076373356259065856 & $3807\pm_{-338}^{+513}$ &$1.2\times10^{33}$ & 5.0 \tabularnewline 
    Gaia DR2 2131725451649506944 & $3190\pm_{-22}^{+27}$ & Gaia DR2 2131725447350716032 & $3602\pm_{-59}^{+62}$ & $2.5\times10^{32}$ & 3.8 \tabularnewline
    Gaia DR2 2132768956904826624 & $3431\pm_{-146}^{+166}$ & Gaia DR2 2132768952604988672 & $4759\pm_{-103}^{+120}$ & $2.9\times10^{32}$ & 26.9 \tabularnewline 
    Gaia DR2 2079073928612819840 & $2776\pm_{-34}^{35}$  & Gaia DR2 2079073928612821760 & $2763\pm_{-33}^{31}$ & $1.7\times10^{31}$ & 1.9 \tabularnewline
    \hline
    \end{tabular}
    \caption{Flare occurrence rate ratios for stars in wide binaries where at least one component had a measured flare occurrence rate. The flare energy is the maximum measured energy for the less active component. For non-equal mass binaries, this is the hotter component.}
    \label{tab:ratios}
\end{table*}

\section{Discussion}

\subsection{Incidence of originally detected flares} \label{sec:incidence}
As stated in Sect\,\ref{sec:results}, \origshiftflares\ flares in our original search came from stars nearby the original \kepler\ target source, or were from multiple flaring stars contributing to the same lightcurve. 
This value comprises $11.7\pm0.5$ per cent of the flares in our original search. This value is similar to the 10-12 per cent false positive rate reported by previous \kepler\ flare studies \citep[which excluded stars with separations less than 12\arcsec, e.g.][]{Shibayama13,Gao16}. We investigated our flare sample further to identify only those flares which came from a neighbouring star. This excludes the flares in wide binaries which come from the target \kepler\ source. We found that 298 flares in our original search came from a neighbouring star (26 stars), whether that was the fainter component of a wide binary or an unrelated nearby source. This new filter reduces the false positive rate to $6.7\pm0.4$ per cent.

Filtering our sample even further to keep only those sources where the nearby star flared (i.e. removing wide binaries in which both components flared) results in a subset of 61 flares from 19 stars. This is $1.4\pm0.2$ per cent of the total number of flares in our original search. In this analysis we have kept flares from non-target stars which have their own entry in the \kepler\ input catalog but lack a distinct short cadence lightcurve.

\subsection{Application to \tess\ flare studies}
\tess\ has since visited the \kepler\ field and studies are now emerging which are combining data from both missions \citep[e.g.][]{Davenport20}. Two of the main differences between the \kepler\ and \tess\ telescopes are the field of view and the pixel scale. The \kepler\ telescope had a total field of view of 115 deg$^2$ and a pixel scale of approximately 4\arcsec, while each camera on the \tess\ telescope has a field of view of 24x24deg$^2$ and a pixel scale of 21\arcsec \citep[][]{Ricker14}. The larger pixel scale of \tess\ will not only cause a greater number of stars to be blended within each aperture, but pairs of stars which were previously resolved with \kepler\ may be placed on the same \tess\ pixel. To confirm whether this was the case for the flare stars in our sample, we calculated both its \tess\ pixel position and the pixel positions of nearby sources in the TESS input catalog. This was done using the Lightkurve Python package \citep[][]{lightkurve18} with the \tess\ sector 14 observations. We found that 21 of the 26 neighbouring flare stars in our sample fell on the same \tess\ pixel as the original \kepler\ target source. The remaining five fell on an adjacent pixel, but all were within 21\arcsec.

In Sect.\,\ref{sec:incidence}, we calculated that \textcolor{black}{$6.7\pm0.4$} per cent of flares in our sample were due to a nearby star either in or neighbouring the pipeline \kepler\ aperture. The smaller pixel size of \kepler\ relative to \tess\ means we can use our survey to estimate the false positive rate of flares in \tess\ flare studies. Specifically, due to the smaller aperture and postage stamp sizes of \kepler\ compared to the \tess\ stamps and full frame images, we can estimate the false positive rate of flares from stars within 21\arcsec, 1 \tess\ pixel distance of the target star. 

Due to the different noise properties between \kepler\ and \tess\, we cannot directly apply our calculated estimate of false positive flare events to large scale flare studies using \tess\ data. In addition, the \tess\ bandpass is not the same as the one used by \kepler\ and it probes a different portion of the flare spectrum. To estimate the number of flares and stars which might cause false positive detections in large scale flare studies using \tess\ short cadence observations, we initially calculated the amplitude of all of our detected flares if they had been observed with the \tess\ bandpass. This combines the flares detected in our original search and those detected after we had created new apertures. To do this for each star we used the best fitting effective temperatures from Sect.\,\ref{sec:sed_fitting} to renormalise a 9000\,K flare blackbody spectrum to the observed amplitude in the \kepler\ bandpass. Both the renormalised flare blackbody spectrum and the PHOENIX v2 spectrum were then convolved with the \tess\ filter to estimate the new amplitude. We then recalculated the dilution from nearby stars in our apertures, using their \tess\ magnitudes obtained from the \tess\ input catalog. The calculated flare amplitude in the \tess\ bandpass was multiplied by the \tess\ dilution coefficient to estimate the flare amplitude in initial \tess\ lightcurves. When calculating the dilution we consider the contributions from stars within a 40 arcsecond radius of the target star \citep[e.g.][]{Tu20}, to better estimate the dilution in \tess\ apertures.

We then used the sample of flares detected with the two minute cadence \tess\ observations from \citet{Gunther2020} to determine the distribution of detectable flares in amplitude-\tess\ magnitude space. We used this distribution to estimate a lower flare amplitude limit as a function of \tess\ magnitude, below which no flares could be detected. We compared this lower limit to our measured flare amplitudes, from our detrended \kepler\ lightcurves, and the \tess\ magnitude of each target. We have used the total sample of flares here, i.e after we revised the apertures to fully include neighbouring stars, to better estimate the larger apertures used for \tess\ relative to \kepler\ \citep[e.g.][]{Tu20}. 
We found that of the 4498 flares detected in our adjusted sample, \textcolor{black}{674} would have an amplitude large enough to be detectable with \tess\ short cadence observations. Of these 674 detectable flares, \textcolor{black}{39} were from stars other than the target source. From this we estimate a false positive rate of \textcolor{black}{$5.8\pm1.0$} per cent for flares from stars within \textcolor{black}{21}\arcsec. This value can be used as an estimate of the rate of false positive events from stars within one pixel as the target in \tess\ flare studies. This value shows that for large scale flare studies using \tess\ short cadence data, a non-negligible fraction of flares will come from stars close to the target source. This effect can be avoided by carefully vetting input catalogs to select only stars with no nearby neighbours.

In this analysis we estimated the dilution in \tess\ by using all stars from within a 40 arcsecond radius. However, many of the \kepler\ image stamps do not extend to 40 arcseconds from the target star, meaning flaring stars at greater separations will have been missed. Consequently, for real \tess\ apertures which extend further than 21 arcseconds from the target source, our estimated false positive rate will be a lower limit. We also note that the false positive fraction will increase in more crowded regions of sky, such as open clusters and those nearer to the galactic plane. As such, caution should be applied when using the quoted values in these regions for studies which have not applied careful vetting of their input catalogues.

\section{Conclusions}
In this work we have presented the results of a search for stellar flares in the \kepler\ short cadence data. We specifically searched for flares which came from stars neighbouring the target sources, which have been excluded from previous surveys. From our search of the short cadence lightcurves using the original \kepler\ apertures we initially detected 4430 flares from 403 stars. Of these, 515 flares came from either a neighbouring source or were from multiple flaring stars contributing to the same lightcurve. We calculated that these flares comprised $11.7\pm0.5$ per cent of the \kepler\ short cadence sample. We found that this value is similar to that reported by previous surveys for the fraction of false positives and blends. Of these 515 flares, 298 were from a neighbouring star. This resulted in a measured false positive rate of $6.7\pm0.4$ per cent in our original sample. We have used our sample and the measured false positive rates to calculate the fraction of flares which would come from sources other than the target star in \tess\ two minute cadence observations. We calculated that $5.8\pm1.0$ per cent rate of flares in these surveys would come from stars within 21\arcsec\ of the target source and recommend careful vetting of input catalogs to minimise the effect of flares from blended sources. 

Our samples and reported false positive rates includes neighbouring stars within 12\arcsec. This is a sample which has been ignored by many previous flare studies using the \kepler\ data. By including these close neighbours in our sample we have been able to study the flaring behaviour of wide binaries, the components of which reside close together on the sky. From our sample of wide binaries, we found that the lower mass components are more active than their higher mass partners during our observations, in agreement with previous studies for the activity of coeval pairs of stars. Our sample highlights how the careful treatment of flares from neighbouring stars can aid in the study of these coeval systems and how flare rates and energies vary between spectral type.   

Where possible we calculated the flare amplitudes and energies for events in our sample, accounting for the flux from both the target source and the flaring neighbour. We found that the flares from neighbouring sources in our sample are some of the highest energy white-light flares observed with \kepler, with energies up to \maxenergyvalue\ erg. We found that the majority of these flares come from M dwarfs which neighbour the \kepler\ target sources. Our study shows how the flares from stars neighbouring the target sources in \kepler, previously discarded as contaminants, can be used to study high energy flare events and \textcolor{black}{flare occurrence rates}. We have focused on the \kepler\ short cadence data in this study, however this data only comprises a small fraction of the total available \kepler\ data. We anticipate that a similar study for the long cadence data will be able to expand upon on the results presented in this study, however will likely be biased to higher energy events due to the 30 minute cadence missing short duration flares.

\section*{Acknowledgements}
JAGJ acknowledges support from grant HST-GO-15955.004-A and R.O.P.L. acknowledges support from grant HST-GO-14784.001-A from the Space Telescope Science Institute, which is operated by the Association of Universities for Research in Astronomy, Inc., under NASA contract NAS 5–26555. 
This paper includes data collected by the Kepler mission. Funding for the Kepler mission is provided by the NASA Science Mission directorate. This publication makes use of data products from the Two Micron All Sky Survey, which is a joint project of the University of Massachusetts and the Infrared Processing and Analysis Center/California Institute of Technology, funded by the National Aeronautics and Space Administration and the National Science Foundation. This work has made use of data from the European Space Agency (ESA) mission
{\it Gaia} (\url{https://www.cosmos.esa.int/gaia}), processed by the {\it Gaia}
Data Processing and Analysis Consortium (DPAC,
\url{https://www.cosmos.esa.int/web/gaia/dpac/consortium}). Funding for the DPAC
has been provided by national institutions, in particular the institutions
participating in the {\it Gaia} Multilateral Agreement. This publication makes use of data products from the Wide-field Infrared Survey Explorer, which is a joint project of the University of California, Los Angeles, and the Jet Propulsion Laboratory/California Institute of Technology, funded by the National Aeronautics and Space Administration. The Pan-STARRS1 Surveys (PS1) and the PS1 public science archive have been made possible through contributions by the Institute for Astronomy, the University of Hawaii, the Pan-STARRS Project Office, the Max-Planck Society and its participating institutes, the Max Planck Institute for Astronomy, Heidelberg and the Max Planck Institute for Extraterrestrial Physics, Garching, The Johns Hopkins University, Durham University, the University of Edinburgh, the Queen's University Belfast, the Harvard-Smithsonian Center for Astrophysics, the Las Cumbres Observatory Global Telescope Network Incorporated, the National Central University of Taiwan, the Space Telescope Science Institute, the National Aeronautics and Space Administration under Grant No. NNX08AR22G issued through the Planetary Science Division of the NASA Science Mission Directorate, the National Science Foundation Grant No. AST-1238877, the University of Maryland, Eotvos Lorand University (ELTE), the Los Alamos National Laboratory, and the Gordon and Betty Moore Foundation. We acknowledge with thanks the variable star observations from the AAVSO International Database contributed by observers worldwide and used in this research.

\section*{Data Availability}
All lightcurves and catalogue broadband photometry used in this work are publicly available. Derived stellar and flare properties are available upon request to JAGJ.

\bibliographystyle{mnras}
\bibliography{references}

\appendix

\bsp	
\label{lastpage}
\end{document}